\documentclass[%
preprint,
superscriptaddress,
 amsmath,amssymb,
 aip,
 pre,
]{revtex4-2}
\usepackage{graphicx}
\usepackage{dcolumn}
\usepackage{bm}
\usepackage{comment}

\usepackage{graphicx}
\usepackage{hyperref}
\usepackage{lineno}

\usepackage{color}
\usepackage[normalem]{ulem}
\def\add#1{\textcolor{black}{#1}}
\def\ds{\displaystyle}
\def\rem#1{}

\def\gs0{\gamma_\mathrm{S0}}
\def\ptl{\partial}
\def\ds{\displaystyle}
\def\ie{\textit{i.e.}, }
\def\eg{\textit{e.g.}, }

\def\bm#1{\mbox{\boldmath $#1$}}
\def\dint{\int\!\!\!\int}
\def\tint{\int\!\!\!\int\!\!\!\int}
\def\rd{\mathrm{d}}
\def\tausf{\tau_\mathrm{sf}}

\def\uslip{u_{\mathrm{slip}}}
\def\eqref#1{(\ref{#1})}
\def\bmfextfs{\bm{f}^\mathrm{ext}_\mathrm{fs}}
\def\bmfextsf{\bm{f}^\mathrm{ext}_\mathrm{sf}}
\def\bmFextsf{\bm{F}^\mathrm{ext}_\mathrm{sf}}
\def\Fextsfx{F^\mathrm{ext}_{\mathrm{sf},x}}
\def\Fextsfy{F^\mathrm{ext}_{\mathrm{sf},y}}
\def\Fextsfz{F^\mathrm{ext}_{\mathrm{sf},z}}
\def\Fextsfa{F^\mathrm{ext}_{\mathrm{sf},\alpha}}
\def\Qf{Q_\mathrm{f}}
\def\Qs{Q_\mathrm{s}}
\def\DQsf{\Delta Q_\mathrm{sf}}
\def\Qfxx{Q_{\mathrm{f},xx}}
\def\Qfyy{Q_{\mathrm{f},yy}}
\def\Qfzz{Q_{\mathrm{f},zz}}
\def\Qfaa{Q_{\mathrm{f},\alpha\alpha}}
\def\rd{\mathrm{d}}

\def\rhof{\rho_\mathrm{f}}
\def\rhos{\rho_\mathrm{s}}
\def\bmuf{\bm{u}_\mathrm{f}}
\def\bmus{\bm{u}_\mathrm{s}}

\def\ufz{u_{\mathrm{f},z}}

\def\ef{e_\mathrm{f}}
\def\es{e_\mathrm{s}}
\def\bmtauf{\bm{\tau}_\mathrm{f}}
\def\bmtaus{\bm{\tau}_\mathrm{s}}
\def\bmJQ{\bm{J}_\mathrm{Q}}
\newcommand{\ifs}{Institute of Fluid Science, Tohoku University, 2-1-1 Katahira, Aoba-ku, Sendai 980-8577, Japan}
\newcommand{\tohoku}{Department of Mechanical Systems Engineering, Tohoku University, 6-6-01 Aramaki Aoba-ku, Sendai 980-8579, Japan}
\newcommand{\osaka}{Department of Mechanical Engineering, the University of Osaka, 2-1 Yamadaoka, Suita 565-0871, Japan}

\begin{document}
\title{
Extraction of slip velocity in NEMD 
Couette flow systems using frictional dissipation}

\author{Hiroki Kusudo}
\thanks{Present affiliation: the University of Osaka}
\email{\add{h.kusudo@mech.eng.osaka-u.ac.jp}}
\affiliation{\tohoku}
\author{Yasutaka Yamaguchi}
\affiliation{\osaka}
\author{Gota Kikugawa}
\affiliation{\ifs}

\begin{abstract}
Velocity slip at the solid--fluid (SF) interface plays a key role in fluid transport at the nanoscale,
and the SF friction coefficient has been extensively studied because it indicates the degree of slippage. 
Owing to the scale of this phenomenon, molecular dynamics (MD) simulations are commonly employed using two major approaches: the Green-Kubo integral method in equilibrium MD (EMD), and the direct calculation of friction force and slip velocity in non-equilibrium MD (NEMD) systems under shear. 
Regarding the latter, a strict definition of the slip velocity is missing due to the nonzero thickness of the boundary at the microscale, and 
the average velocity of the first adsorption layer 
or the velocity at the boundary obtained by extrapolation or interpolation is often used. 
In this study, we propose an alternative description of the slip velocity based on a thermal perspective from the two different scales, i.e.,  
at the macroscale, frictional heat is defined as the product of the friction force and slip velocity, 
whereas at the microscale, 
it can be expressed as the sum of the works exerted on the fluid and solid by each other.
By combining the two different scales, 
we defined the slip velocity based on the dissipation induced at the SF interface under shear, 
which avoids the arbitrariness in the slip velocity at the microscale.
\end{abstract}

\maketitle
%
%
%
\section{Introduction}
Recently, understanding fluid behaviors at the micro- and nanoscales is being increasingly required 
with the development of nanofluidic devices and  nanotechnology.~\cite{eijkel_nanofluidics_2005, bocquet_nanofluidics_2010}
In fluid transport at the nanoscale, 
the fluid slip at the interface plays a key role,~\cite{falk_molecular_2010, bocquet_nanofluidics_2010, bocquet_flow_2007, thomas_pressure-driven_2010, sam_water_2019, kannam_interfacial_2012, kannam_how_2013, falk_molecular_2010}
which is characterized by the difference of the tangential velocity components of the solid and fluid (SF) at the interface.
This is in contrast to macroscopic fluid dynamics where 
a non-slip boundary condition (BC) is usually imposed on solid surfaces; excluding the field of rarefied gas dynamics considering the Knudsen layer. 
The historical discussion up to the first half of the 20th century on this topic is summarized in Goldstein's textbook.~\cite{Goldstein1938}
Moreover, regarding the solid--liquid (SL) interfaces, 
recent developments in nanoscale measurement techniques and computational science, 
such as molecular dynamics (MD), have revealed the existence of a non-zero velocity slip.
\par
As a macroscopic BC for the slip at a SF interface, 
Navier~\cite{navier_memoire_1823} suggested the following relation:
\begin{equation}
    \tausf=\lambda \uslip,
    \label{eq:Navier}
\end{equation}
where $\tausf$ and $\uslip$ are the friction force per unit area and the slip velocity, and the two are linearly correlated with $\lambda$ called the friction coefficient (FC). 
Equation~\eqref{eq:Navier} is known as 
the Navier BC while 
the slip length is often used to characterize the slip
as well, which is defined as the distance from the SF interface to the position at which the fluid velocity profile extrapolated to the solid side corresponds to the solid velocity.
This is clear from a macroscopic viewpoint with an assumption that the interface is a zero-thickness boundary between homogeneous fluid and solid. 
%
%
%
%
%
%
If this homogeneous fluid obeys Newton's law of viscosity, the shear force per area, \ie shear stress, is given by 
\begin{equation}\label{eq:newtonvisc}
    \tausf = \mu \left. \frac{\ptl u}{\ptl z} \right|_\text{interface},
\end{equation}
where $\mu$ and $u$ are the fluid viscosity and the velocity parallel to the SF interface as a function of the interface-normal position $z$.
Then, the Navier BC in Eq.~\eqref{eq:Navier} is written in another form as
\begin{equation}
    \frac{\uslip}{b} 
    = 
    \left. \frac{\ptl u}{\ptl z}\right|_{\text{interface}},
\end{equation}
where $b$ is called the slip length given by
\begin{equation}\label{eq:def_b}
 b  =\frac{\mu}{\lambda}.
\end{equation}
%
The slip length depends on the SF combination; \eg for water on a graphene or carbon nanotube surface, $b$ was estimated to be about several tens of nanometers from experiments and MD simulations.~\cite{falk_molecular_2010, Keerthi2021, Chen2021} 
%
\par
Several methods have been proposed to obtain the FC or the slip length for SL interfaces based on 
both non-equilibrium MD (NEMD) 
~\cite{kannam_interfacial_2012, hansen_prediction_2011, herrero_shear_2019, liu_molecular_2023, mundy_hydrodynamic_1996, shi_study_2023, herrero_fast_2020, nakaoka_molecular_2017, huang_friction_2012, omori_full_2019, maffioli_slip_2022, martini_slip_2008, sam_water_2019, oga_green-kubo_2019}
and equilibrium MD (EMD).~\cite{bocquet_hydrodynamic_1994, varghese_improved_2021, hadjiconstantinou_equivalence_2022, bui_revisiting_2024, ramos-alvarado_hydrodynamic_2016, oga_green-kubo_2019, oga_equilibrium_2023, hansen_prediction_2011, 
KumarKannam_EMD_2012, Petravic2007, nakano_microscopic_2019, Bocquet2013, Huang2014, Sam2018, Sokhan2008, Kiefer2025}
%
%
%
%
For the former, a Couette-type flow is often adopted using a system with a liquid confined between two parallel solid walls, where the walls are moving at a constant velocity toward the opposite directions tangential to the interface, and $\lambda$ is calculated by Eq.~\eqref{eq:Navier} using the resulting $\tausf$ and $\uslip$ as the time average in the steady-state.
This seems rather simple; however, two fundamental 
problems arise with the definition of the slip velocity $\uslip$. 
The first is about the velocity profile because the liquid molecules at the solid interface form an anisotropic layered structure, which apparently has a viscosity and density different from those in the homogeneous bulk.
The other is about the strict definition of the SL 
interface position because the SL interface has a 
finite thickness typically having a length scale of the diameter of solid and fluid molecules.
Regarding the first, the velocity profile in the liquid bulk away from the interface is often used for the extrapolation,~\cite{bocquet_flow_2007, liu_molecular_2023} 
while even with the extrapolation, the slip length $b$
cannot be determined without a strict definition of 
the SL interface position. 
%
Another possibility is directly using the average velocity of the first liquid adsorption layer on the solid surface to determine $\uslip$. Considering that more than 80~\%
of the SL shear force from the solid is exerted on the first layer for a flat crystal surface,~\cite{Qian2003}
this seems reasonable from the original idea of Eq.~\eqref{eq:Navier}; however, the correspondence with the macroscopic feature is not clear.~\cite{bocquet_hydrodynamic_1994, hadjiconstantinou_equivalence_2022, nakano_microscopic_2019, mundy_hydrodynamic_1996, camargo_boundary_2019}
\par
Some studies tried to identify the slip velocity using a similar system with a liquid confined between two parallel solid walls. \citet{herrero_shear_2019} measured the shear force on the solid walls in a Poiseuille-type flow system driven by exerting \add{a} gravity-like constant external force on the fluid molecules and evaluated the hydrodynamic height, \ie the interface position by assuming a parabolic velocity distribution. In the MD simulations of \citet{omori_full_2019}, 
the frequency dependence of the SL friction was examined by simultaneously oscillating the walls parallel to the interface at 
various frequencies.
%

On the other hand, several methods have also been proposed to examine the SF friction based on the Green-Kubo (GK) relation using EMD systems, 
where the SL systems were at static equilibrium without shear or external driving force and the autocorrelation and/or the cross-correlation of the force 
or velocity were analyzed.~\cite{KumarKannam_EMD_2012, bocquet_hydrodynamic_1994, Petravic2007, nakano_microscopic_2019, hadjiconstantinou_equivalence_2022, hansen_prediction_2011, Bocquet2013, Huang2014, Sam2018, oga_green-kubo_2019, varghese_improved_2021, Sokhan2008, Kiefer2025, Espanol2019}
%
It is known that the integral of the autocorrelation function of 
the SL friction force, called the GK integral, 
shows a distinctive difference from that in bulk systems; for instance 
in a bulk liquid system without interface, the integral of 
the shear stress autocorrelation monotonically increases with time 
and converges to a constant value corresponding to the 
shear viscosity, whereas the GK integral
of the SL friction force usually increases with time and decays 
after taking its maximum. This difference is called the plateau problem,~\cite{Espanol2019}
and recent studies have explained the mechanism of this problem through 
the derivation of the solutions of the GK integral based on the Stokes equation, 
and system-size dependent behavior of the 
GK integral was shown as well.~\cite{oga_equilibrium_2023, nakano_microscopic_2019, Kiefer2025}
\par
%
%
%
\par
In this study, we took notice of the dissipative process due to 
the SF friction leading to heat generation in order to investigate the
velocity slip from another viewpoint of energy conservation. More concretely, 
in a steady-state Couette flow system, the energy conservation indicates that 
the divergence of the heat flux, \ie the generated heat in the bulk having a 
finite volume, 
balances the divergence of the stress work, \ie the viscous dissipation heat.~\cite{khare_molecular_2006, chen_molecular_2017, han_method_2004, zhang_pressure_2004, kusudo_receding_2023}
We extend this scenario to a SF interface under shear, where 
the frictional heat generated at a SF interface induces a discontinuous heat flux there. 
%
%
We related this heat flux discontinuity with the heat 
dissipation by the SF friction force to determine the 
slip velocity by assuming that the heat dissipation is 
given by the product of the SF friction force and the 
slip velocity.
%
%
%
%
We adopted this methodology for NEMD simulations of a 
quasi-one-dimensional (1D) Couette-type flow system of a 
Lennard-Jones liquid confined between parallel crystal walls 
with various wettability, and compared the results 
with those obtained by the GK-based method in our previous 
study as a reference. 
As an advantage, the slip velocity can be determined 
by this method without the need of physical properties 
including viscosity and without assuming the interface 
position or the extrapolation of the velocity profile.
%
%
\par
%
%
%
%
\section{Method}
\subsection{Theoretical Framework}
\begin{figure}
  \begin{center}
    \includegraphics[width=.9\linewidth]{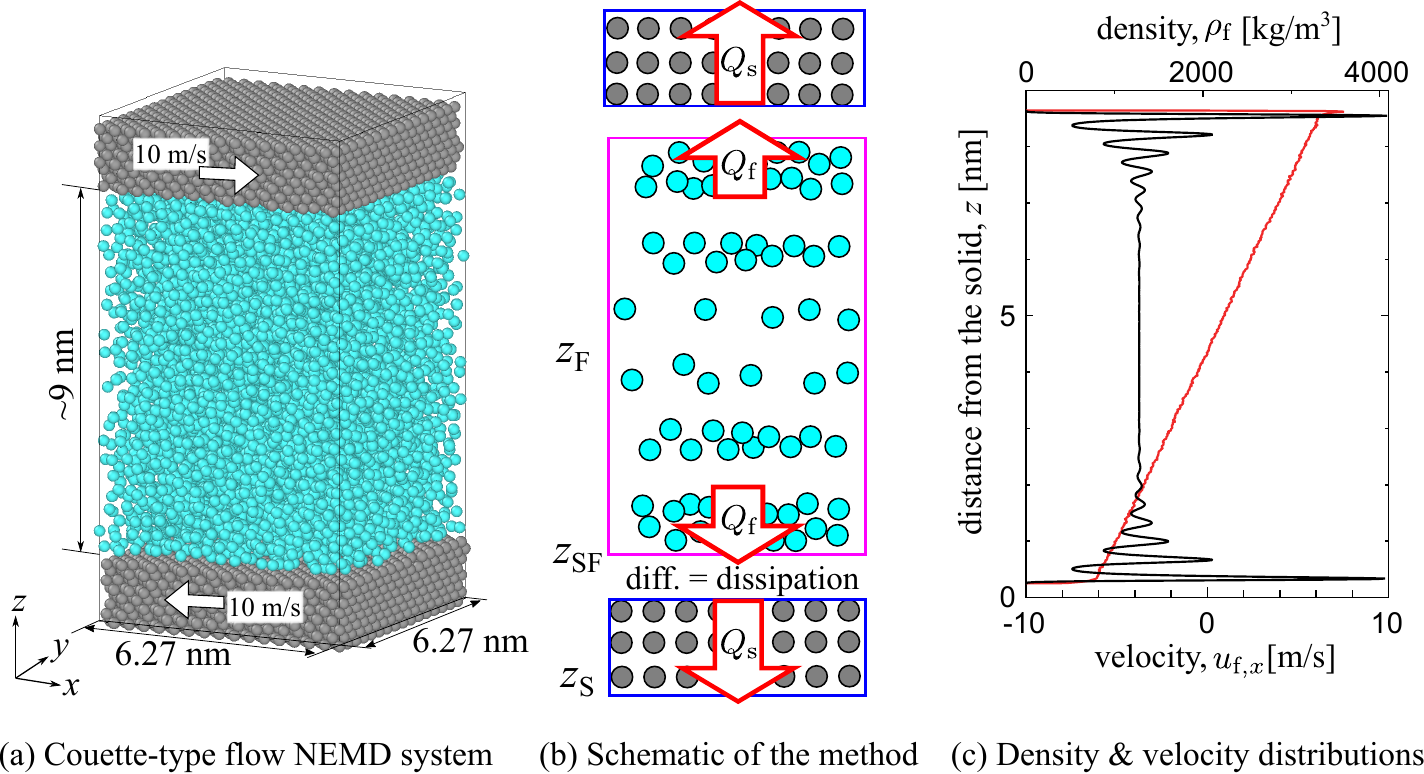}
  \end{center} 
  \caption{ \add{(a)} Couette-type flow NEMD system used in this study. \add{(b)} Schematic of the present methodology. \add{(c)} Distributions of the fluid density \add{(black line) }and velocity \add{(red line) }with a wetting parameter of $\eta=0.5$, which corresponds to an equilibrium droplet contact angle of $46^\circ$.
  \label{fig:fig1} 
}
\end{figure}
In this study, we extracted the slip velocity at a SL interface based on the dissipation due to friction in a Couette-type flow NEMD system exemplified in Fig.~\ref{fig:fig1}. While the details of the calculation are described later in Sec.~\ref{sec:system}, we provide the basic recipe through the evaluation of 
the heat flows into the solid and out of the fluid through the SL interface in such a system here.
Considering the direction of the heat flow, we denote the heat gain of the solid 
and the heat loss of the fluid 
per unit area as $\Qs$ and $\Qf$, respectively. 
Note that both are positive as in Fig.~\ref{fig:fig1}(b). Then, 
the difference $\DQsf$ defined by
\begin{align}
    \DQsf \equiv \Qs - \Qf 
    \label{eq:deltaJ}
\end{align}
accounts for the dissipation at the interface
due to friction.
In the macroscopic description, 
it corresponds to the frictional heat expressed by
\begin{equation}
\DQsf = \tausf \uslip, 
\label{eq:DeltaQ=tau_uslip}
\end{equation}
where $\tausf$ and $\uslip$ denote the friction force per unit area and the slip velocity, respectively.
Then, the slip velocity $\uslip$ can be written as
\begin{equation}
\uslip=\frac{\DQsf}{\tausf}=\frac{\Qs - \Qf}{\tausf}. 
    \label{eq:uslip}
\end{equation}
This shows that, in the macroscopic description, the slip velocity can be obtained 
by calculating the friction force $\tausf$ and the difference 
of the heat gain of the solid and the heat loss of the fluid. 
\par
In the following, we derive the heat flux difference $\DQsf$ as the friction-induced dissipation 
using a \add{macroscopic formulation assuming the following conditions to be applied to MD systems: 
\begin{itemize}
    \item The system is in a steady state and one-dimensional flow.
    \item The fluid--fluid interaction is treated as fluid stress.     
    \item The solid--fluid interaction is treated as volumetric external force. 
    \item The two solid walls are rigid and move at a constant velocity in opposite directions parallel to the wall.
\end{itemize}
}
We start from the energy conservation of a fluid with an external force expressed by
\begin{align}
\label{eq:enecon}
\frac{\partial \rhof \ef}{\partial t} = -\nabla \cdot  
( \rhof \ef \bmuf + \bm{J}_{\text{Q}}-\bmtauf \cdot \bmuf)
+
\rho \bmfextsf\cdot \bmuf,
\end{align}
where $\rhof$, $\ef$ and $\bmuf$ denote the fluid density, specific energy, and velocity, respectively, while
$\bmJQ$, $\bmtauf$ and $\bmfextsf$ denote the heat flux, the stress tensor in the fluid, and the external body force per unit mass exerted on the fluid by the solid, respectively.
Note that the molecular-scale description of the stress is composed of the momentum transfer due to the molecular motion and the impulse of the interaction force,~\cite{Todd_Daivis2017} and 
also note that in this study, the SF interaction is treated as an external body force, whereas the fluid-fluid interaction is included in the stress tensor as in our previous studies.~\cite{kusudo_local_2021,kusudo_receding_2023} 
We transform Eq.~\eqref{eq:enecon} as
\begin{align}
\label{eq:divHF}
\nabla \cdot \bmJQ
=-\frac{\partial \rhof \ef}{\partial t} 
-\nabla \cdot  \left( \rhof \ef \bmuf \right)
+\nabla \cdot  \left(\bmtauf \cdot \bmuf \right)+
\rho \bmfextsf \cdot \bmuf.
\end{align}
Equation~\eqref{eq:divHF} means that the divergence of the heat flux 
in the left-hand side (LHS) can be obtained by calculating the right-hand side (RHS)
composed of the time derivative of the fluid energy, 
the energy advection term, the stress work term, 
and the work done by the external force from the solid.
We applied this framework to a steady-state quasi-1D system, \ie the Couette-type flow system shown in Fig.~\ref{fig:fig1}. 
Then, the first and second terms on the RHS of Eq.~\eqref{eq:divHF} are zero, which yields
\begin{align}
\label{eq:divHF_Couette}
\nabla \cdot \bmJQ
=\nabla \cdot \left( \bmtauf \cdot \bmuf \right) +
\rhof \bmfextsf \cdot \bmuf.
\end{align}
Note that in the case of the Couette-type flow system, the first term on the RHS corresponds to the energy dissipation in the liquid phase due to viscosity. 
By integrating Eq.~\eqref{eq:divHF_Couette} and by applying Gauss'
divergence theorem for an arbitrary control volume (CV) of volume $V$ surrounded 
by closed surface $S$, it follows
%
\begin{align}
 \dint_{S} \rd \bm{S} \cdot 
    \bm{J}_{\text{Q}}
    =
    \dint_{S} \rd \bm{S} \cdot 
    \left( \bmtauf \cdot \bmuf \right) +
    \tint_{V} \rd  V 
    \rhof \bmfextsf \cdot \bmuf,
\label{eq:divHF_int}
\end{align}
%
where $\rd \bm{S}\equiv \bm{n} \rd S$ and $\rd V$ denote the infinitesimal 
surface element of $S$ with the unit normal $\bm{n}$ outward 
$S$ and infinitesimal volume element of $V$. Equation~\eqref{eq:divHF_int} 
means that the outward heat flow from the CV through $S$ 
in the LHS 
can be obtained by calculating the mechanical works due to the stress across the surface $S$ 
and due to the external body force exerted on the fluid by the solid
in the RHS. 
Let us consider the CV set to surround all fluids but exclude the solid as indicated by the magenta rectangle in Fig.~\ref{fig:fig1}(b). 
Under the periodic boundary conditions in the lateral $x$- and $y$-directions, 
the stress work as the first term on the RHS of Eq.~\eqref{eq:divHF_int} becomes zero because neither fluid passage nor fluid-fluid interaction 
crosses the CV surface parallel to the solid surface.~\cite{yamaguchi_interpretation_2019,kusudo_local_2021}
Considering the quasi-one-dimensional feature of the system and symmetry in the surface-normal direction, 
the fluid heat loss due to the solid per unit area can be obtained simply from the friction work on the fluid in the CV as
\begin{align}
\Qf
=
\int_{z_{\text{SF}}}^{z_{\text{F}}} \rd z\: 
    \rho \bmfextsf \cdot \bmuf.
\label{eq:JQFW2}
\end{align}
In Eq.~\eqref{eq:JQFW2}, $z_{\text{SF}}$ denotes the vertical position of the bottom face of the CV and
$z_{\text{F}}$ denotes a position sufficiently far from the solid surface where the solid-fluid interaction is negligible, \ie out of the cutoff distance of the solid-fluid interaction.
\par
On the other hand for the solid, 
similar to the fluid, the heat gain of the solid due to the fluid can be derived. 
The energy conservation of the solid is expressed by
\begin{align}
\label{eq:eneconw}
\frac{\partial \rhos \es}{\partial t} 
= 
-\nabla \cdot  ( \rhos \es \bmus + \bmJQ -\bmtaus \cdot \bmus) \nonumber \\
+
\rhos \bmfextfs \cdot \bmus
-\dot{Q}_{\rm{thermo}}+\bm{F}_{\rm{drive}}\cdot \bmus,
\end{align}
where $\rhos$, $\es$ and $\bmtaus$ denote the solid density, the specific energy, and the stress in the solid, respectively. 
Note that $\bmfextfs$ is an external body force exerted on the solid by the fluid
opposite to $\bmfextsf$.
Also note that $\dot{Q}_{\rm{thermo}}$ is the heat removal due to thermostat and $\bm{F}_{\rm{drive}}$ is the driving force applied outside of the system for moving the solid layers.
Similar to Eq.~\eqref{eq:divHF_Couette} of the fluid side, for the steady-state quasi-1D system, Eq.~\eqref{eq:eneconw} is transformed as 
\begin{align}
\label{eq:divHF_Couette_s}
\nabla \cdot \bmJQ
+\dot{Q}_{\rm{thermo}}
=\nabla \cdot \left( \bmtaus \cdot \bmus \right) +
\rhos \bmfextfs \cdot \bmus
+\bm{F}_{\rm{drive}}\cdot \bmus.
\end{align}
Note that $\bmJQ$ denotes the heat flux inside the system.\rem{ whereas $\dot{Q}_{\rm{thermo}}$ denotes the heat removal from the system due to thermostat.}
Again, we set a CV to surround only the solid as well 
indicated by the blue rectangles in Fig.~\ref{fig:fig1}(b). 
By integrating Eq.~\eqref{eq:divHF_Couette_s} for the CV, the total outward heat flow from the CV can be obtained, which includes the heat loss due to the temperature control $\dot{Q}_{\rm{thermo}}$ imposed on the wall. 
Such a heat loss of the second term on the LHS balances the work for moving wall of the third term on the RHS because the system is in a steady state. 
In addition, the stress work as the first term in the
RHS of Eq.~\eqref{eq:divHF_Couette_s} becomes zero because neither solid passage nor solid-solid interaction crosses
the CV surface parallel to the solid surface.
%
Then, the solid heat gain $\Qs$ through the SL interface per unit area, \ie the heat loss with its sign inverted, can be obtained as
\begin{align}
\Qs = -
\int_{z_{\text{S}}}^{z_{\text{SF}}} \rd z\:
    \rhos \bmfextfs \cdot \bmus.
\label{eq:JQWF}
\end{align}
In Eq.~\eqref{eq:JQWF}, $z_{\text{S}}$ denotes a position sufficiently far from the fluid where the solid-fluid interaction is negligible, \ie out of the cutoff distance of the solid-fluid interaction.
In addition, we define the total external force \add{per unit area} $\bmFextsf$ exerted on the fluid by the solid as the volume integral of the local external force $\bmfextsf$, as follows:
\begin{equation}
    \bmFextsf 
    \equiv
    \int_{z_{\text{SF}}}^{z_{\text{F}}} \rd z\: 
    \rhof \bmfextsf
    =
    -\int_{z_{\text{S}}}^{z_{\text{SF}}} \rd z\:
    \rhos \bmfextfs,
    \label{eq:frictionforce}
\end{equation}
%
where the second equality is satisfied because of the action-reaction relation.
Note that the component of $\bmFextsf$ in the shear direction, 
\ie the force in the $x$-direction for the bottom interface 
in Fig.~\ref{fig:fig1}, corresponds to the friction force \add{per unit area} exerted on the solid surface with the opposite sign
$-\tausf$.
Since the solid velocity $\bmus$ is uniform,  
Eq.~\eqref{eq:JQWF} is transformed into 
\begin{align}
\Qs = \bmFextsf \cdot \bmus.
\label{eq:JQWF2}
\end{align}
Finally, by substituting Eqs.~\eqref{eq:JQFW2} and \eqref{eq:JQWF2} 
into Eq.~\eqref{eq:uslip}, the macroscopic slip velocity is written as
\begin{align}
  \uslip
  =
  \frac{\Qs - \Qf}{\tausf}
  =
  \frac{
  \bmFextsf \cdot \bmus
  -
  \ds \int_{z_{\text{SF}}}^{z_{\text{F}}} \rd z\:
 \rho \bmfextsf \cdot \bmuf
 }{\tausf}
.
  \label{eq:u_slip1D}
\end{align}
This means that the slip velocity is defined based on the integration of velocity $\bmuf$ weighted by the friction force \add{per unit volume} $\rhof \bmfextsf$.
Note that $\uslip$ and $\tausf$ have the same sign and can be both positive and negative.
Furthermore, by applying the Navier boundary condition in Eq.~\eqref{eq:Navier}, 
the FC can be determined 
by calculating only the distributions of the fluid velocity and external body force, \ie the fluid-solid interaction force per unit volume, as follows:
\begin{align}
  \lambda
  &=
  \frac
  {\tausf^{2}}
  {
  \bmFextsf \cdot \bmus
  -
    \ds \int_{z_{\text{SF}}}^{z_{\text{F}}} \rd z\:
    \rhof \bmfextsf \cdot \bmuf}.
  \label{eq:beta1D}
\end{align}
The denominator indicates again that the solid has a uniform velocity $\bmus$
whereas the fluid has non-uniform velocity $\bmuf$; this velocity difference between solid and fluid gives rise to
heat dissipation due to friction.
\par
In the following section, we obtained the slip velocity $\uslip$
by Eq.~\eqref{eq:u_slip1D} and compared with the velocity 
distributions of the fluid, and also calculated the friction 
coefficient $\lambda$ by Eq.~\eqref{eq:beta1D} and 
compared with the results in our previous study.~\cite{oga_equilibrium_2023}
\par
%
%
%
%
\subsection{MD Simulation System}
\label{sec:system}
All the simulations were performed using the LAMMPS package (29Sep2021).~\cite{thompson_lammps_2022} 
In this study, we demonstrated the proposed 
methods by performing NEMD simulations 
of a quasi-1D Couette flow system, which is shown in Fig.~\ref{fig:fig1}(a), for various wettability cases. 
The basic setup was the same as that used in our previous study,~\cite{oga_equilibrium_2023} 
and the values are summarized in Table~\ref{tab:table1} with the non-dimensional units normalized by the corresponding standard values based on $\sigma_{\text{ff}}$, $\epsilon_{\text{ff}}$ and $m_{\text{f}}$.
The fluid--fluid and fluid--solid interactions were modeled using the 12-6 Lennard-Jones(LJ) potential 
$\Phi^\mathrm{LJ}(r_{ij}) =  4\epsilon_{ij} \left[ \left(\frac{\sigma_{ij}}{r_{ij}}\right)^{12}-
\left(\frac{\sigma_{ij}}{r_{ij}}\right)^{6} \right]$, 
where $r_{ij}$ is the distance between particles $i$ and $j$ and $\epsilon_{ij}$ and $\sigma_{ij}$ denote the LJ energy and length parameters, respectively.
Quadratic functions were added to this LJ potential, such that the potential and interaction force smoothly vanished at a cut-off distance of $r_\mathrm{c}=3.5 \sigma$.~\cite{Nishida2014}
We used the following fixed parameters for the fluid--fluid (ff) and fluid--solid (fs) interactions:
$\sigma_{\text{ff}}=0.340$\,nm, $\epsilon_{\text{ff}}=1.67\times10^{-21}$\,J, 
$\sigma_{\text{fs}}=0.345$\,nm, 
whereas $\epsilon_\mathrm{sf} = \eta \epsilon^{0}_\mathrm{sf}$ 
was changed by multiplying the base value 
$\epsilon^{0}_\mathrm{sf}=1.29\times10^{-21}$~J 
by the wetting parameter $\eta$ 
ranging from $\eta=0.2$ to $0.6$ in the present simulations
considering the comparison with our previous study.~\cite{oga_equilibrium_2023}
The equilibrium droplet contact angles are approximately $138^\circ$ for $\eta=0.2$, $46^\circ$ for $\eta=0.5$, and complete wetting for $\eta=0.6$ at the present control temperature of 100~K.
\add{The solid--solid interaction was modeled by a harmonic spring with equilibrium distance of $r_{0}=0.277$~nm and spring constant of $k_0=46.8$~N/m, so that the Young's modulus was around 200~GPa and the solid walls can be regarded as effectively rigid in the present simulation.}
The atomic masses of the fluid and solid particles were $m_{\text{f}}=39.95$\,u and $m_{\text{s}}=195.1$\,u, respectively.
Finally, the equations of motion were integrated using the velocity-Verlet algorithm with a time step $\Delta t$ of 5\,fs.
\par
Periodic boundary conditions were set in the $x$- and $y$-directions, and 6400 LJ particles were 
confined between two parallel solid walls (dimensions of $x \times y = 6.27\times 6.27$\,nm$^{2}$) at a distance of $\sim 9$\,nm.
To achieve a steady-state system, temperature control was applied to the second outermost layers of the solid walls using a Langevin thermostat at $T_{\mathrm{s}}=100$\,K, 
to remove the viscous heat in the liquid under shear.
%
\par
The system was equilibrated for 10~ns \add{by applying an external force in the $z$-direction, equivalent to the target pressure of 4~MPa, to} the outermost layer of the top wall. 
The distance between the walls was determined based on the time-averaged position of the top wall 
during a further pre-calculation of 5~ns.
After equilibration, \add{the top wall was constrained in the $z$-direction, and}
to achieve a steady shear flow, 
another relaxation run was performed 
for 5~ns by moving the particles in the outermost layers of both walls at opposite velocities of $\pm$10~m/s in the $x$-direction. 
\add{The resulting pressure, defined as the time-averaged force per unit area exerted on the bottom wall by the fluid in the $z$-direction,} was approximately equal to the control pressure of 4~MPa. This is shown in Fig.~\ref{fig:fig3}(a).
%
Finally, the main calculation was performed for an effective averaging time of 200\,ns, obtained from 40 independent block averages over 5\,ns. 
We calculated the volume-averaged solid--fluid interaction force, \ie the external body force, fluid density, and fluid velocity 
using flat bins normal to the $z$-direction with a thickness 
of $\Delta z = 0.0102$~nm. 
In this study, we calculate the fluid density, velocity and friction force distribution by the volume average method to obtain the slip velocity by Eq.~\eqref{eq:u_slip1D}. 
Each macroscopic quantity averaged for the local bin of volume $\Delta V$ is calculated as follows
\begin{align}
    \rhof&=\frac{ \left< \sum\limits_{i \in \text{fluid}}^{\text{in} \Delta V} m_{i} \right>}{\Delta V} 
    \\
    \rhof \bmuf&=\frac{ \left< \sum\limits_{i \in \text{fluid}}^{\text{in} \Delta V} m_{i}\bm{v}_i \right>}{\Delta V}
    \\
    \rhof \bmfextsf&=\frac{ \left< 
    \sum\limits_{i \in \mathrm{fluid}}^{\text{in} \Delta V} 
    \sum\limits_{j \in \mathrm{solid}} 
    \bm{F}^{ij}
    \right>}{\Delta V},
\end{align}
where $\bm{v}_{i}$ and $\bm{F}^{ij}$ denote the microscopic velocity of particle $i$ and the interaction force vector exerted on particle $i$ from particle $j$, respectively and $\left< \right>$ denotes the ensemble average, which can be replaced by the time average because the present system is in a steady state and considered ergodic. 
Additionally, the volume-averaged velocity, which is used in Eq.~\eqref{eq:u_slip1D}, can be calculated as 
\begin{align}
    \bmuf=\frac{\rhof \bmuf}{\rhof} 
    =\frac{ \left< \sum\limits_{i \in \text{fluid}}^{\text{in} \Delta V} m_{i}\bm{v}_i \right>}{\left< \sum\limits_{i \in \text{fluid}}^{\text{in} \Delta V} m_{i} \right>}.
\end{align}

\begin{table*}[!t]
\caption{\label{tab:table1} 
Simulation parameters and their corresponding non-dimensional values.
}
\begin{ruledtabular}
\begin{tabular}{cccc}
property  & value & unit & non-dim. value
\\ \hline
$\sigma_\mathrm{ff}$ & 0.340 & nm & 1
\\
$\varepsilon_\mathrm{ff}$ & $1.67 \times 10^{-21}$ & J & 1
\\
$\sigma_\mathrm{sf}$ & 0.345 & nm & 0.772
\\
$\varepsilon^{0}_\mathrm{sf}$
& $1.29\times 10^{-21}$ & J & 1.18
\\
$\varepsilon_\mathrm{sf}$ & 
$\eta \times \varepsilon^{0}_\mathrm{sf}$
\\
$\eta$ &
0.2 -- 0.6 & - & -
\\
$m_\mathrm{f}$ & $6.64 \times 10^{-26}$ & kg & 1
\\
$m_\mathrm{s}$ & $3.24 \times 10^{-25}$ & kg & 4.89
\\
\add{$r_0$} & \add{0.277} & \add{nm} & \add{0.815} 
\\
\add{$k$} & \add{46.8} & \add{N/m} & \add{$3.24\times 10^3$}
\\
$T$ & 100  & K & 0.827
\\
$N_\mathrm{f}$  & 6400  & - & -
\\
\end{tabular}
\end{ruledtabular}
\end{table*}
\par
%
%
%
%
\section{Results and Discussion}
As a representative case of a Couette-type flow in the present NEMD simulations, 
distributions of the density and velocity for a wetting parameter of $\eta=0.5$ are shown in Fig.~\ref{fig:fig1}(c). 
A clear difference is observed between the velocities of the solid ($\pm 10$~m/s) and fluid at the solid interface. In addition, the velocity profile is not on a straight line, \ie the velocity gradient is not constant,  near the solid, where a clear layered structure is observed. 
Hereafter, we focus on the region in the vicinity of the solid, specifically within 1~nm, \ie within the strong interaction range of the solid--fluid force given by the short-range LJ potential.
\begin{figure}[b]
  \begin{center}
    \includegraphics[width=\linewidth]{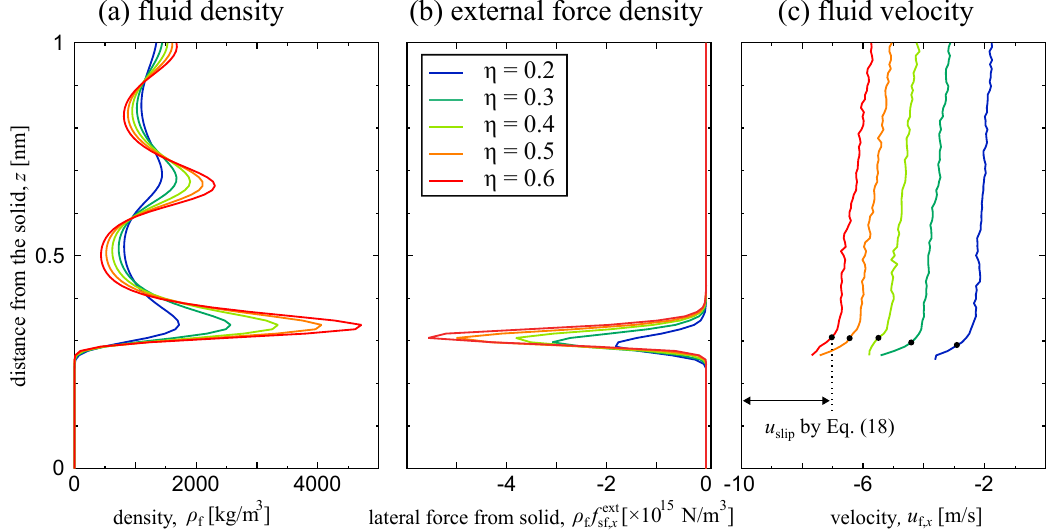}
  \end{center} 
  \caption{\label{fig:fig2} 
  Distributions of the (a) fluid density, (b) external force density exerted by the solid wall, and (c) fluid velocity for various wetting parameters $\eta$. Corresponding points to the slip velocity $\uslip$ obtained by Eq.~\eqref{eq:u_slip1D} on the velocity distributions are displayed in (c).
}
\end{figure}
\par
Distributions of the fluid density $\rhof$, the shear force density 
$\rhof f^{\mathrm{ext}}_{\mathrm{sf},x}$ 
exerted on the fluid by the solid as an external force, and the velocity are shown in Fig.~\ref{fig:fig2} for various wetting parameters $\eta$ between $0.2$ and $0.6$, where the origin of the vertical axis $z$ is set at the average position of the surface layer of the bottom fcc solid wall, and 
$f^{\mathrm{ext}}_{\mathrm{sf},x}$
is defined by
\begin{equation}
f^{\mathrm{ext}}_{\mathrm{sf},x}
\equiv
\bmfextsf\cdot \bm{n}_{x}
\label{eq:def_fextsfx}
\end{equation}
with $\bm{n}_{x}$ being the unit vector in the $x$-direction.
In addition, we obtained the slip velocity by Eq.~\eqref{eq:u_slip1D}, 
and estimated the corresponding position, indicated by black points, on the velocity distributions in Fig.~\ref{fig:fig2}(c). 
These points are considered to be the positions at which the slip velocity should be evaluated
\add{ and are located around the peak position of the external force density in Fig.~\ref{fig:fig2}(b)}. 
\add{This coincides with the Navier BC in Eq.~\eqref{eq:Navier}, 
because friction force is exerted as a surface force at the boundary.}
As seen in the density distribution in Fig.~\ref{fig:fig2}(a), 
the fluid formed multiple adsorption layers near the solid surface, 
and the density of each layer increased with the wetting parameter.
On the other hand, most of the friction force was exerted on the first 
adsorption layer closest to the solid side as observed in 
Fig.~\ref{fig:fig2}(b) especially in the present system with a short-range solid-fluid interaction.~\cite{Qian2003} 
As the first adsorption density increased with the wetting parameter, the friction force also increased.
This was not only due to the large amount of the fluid molecules interacting with the solid, but also influenced by the wetting parameter.
In addition, it is worth noting that the peak positions of the density of the first layer and the external force density is slightly different, meaning that the friction force on the solid was mainly exerted not from the liquid molecules at the equilibrium position of the liquid giving the highest density but those at positions slightly closer to the 
solid subject to repulsive interaction with the solid.
%
Regarding the slip velocity in Fig.~\ref{fig:fig2}(c), it decreased with 
increasing the wetting parameter as intuitively expected.~\cite{Huang2008}
Note that the velocity is displayed only in the region where the density is larger than 10~$\text{kg/}\text{m}^3$.
In addition, the 
positions to evaluate the slip velocity 
corresponded rather the peak positions of the external force density than those of the liquid density, and the slip position was closer to the solid surface for lower wettability case of 
smaller $\eta$.
\add{This can be understood by considering the SL interaction force to be mainly due to repulsive interaction at short distance. 
At lower wettability in the present system, the closest approach distance of the fluid particles to the solid surface, which gives strong repulsive interaction,  decreases, 
resulting in the decrease in the average distance.
Therefore, the peak of the resulting force density shifts toward the solid surface with lower wettability, 
as shown in Fig.~\ref{fig:fig2}(b).} 
%
At the microscale, the definition of the fluid velocity associated with the slip velocity is not unique.
To avoid arbitrariness in the definition of slip velocity, 
in this study, 
we deal with 
the heat dissipation induced at the friction interface.
%
\par
\begin{figure}[tb]
  \begin{center}
    \includegraphics[width=\linewidth]{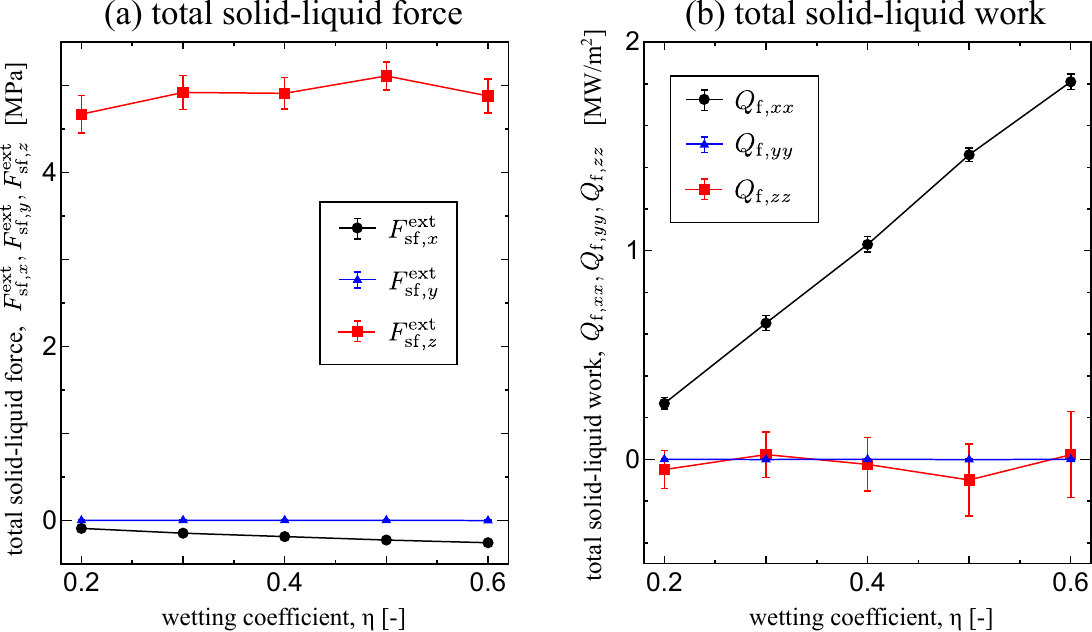}
  \end{center} 
  \caption{\label{fig:fig3} 
  (a) Directional components of the total force on the liquid by the 
  solid, and (b) their contributions to the solid-liquid work
  for various wetting parameters $\eta$. 
  Both are per unit area as the definitions given in Eqs.~\eqref{eq:Fextsf_components} and \eqref{eq:frictionwork},
  respectively.
}
\end{figure}
We used Eq.~\eqref{eq:u_slip1D} to extract the slip velocity which 
includes the external work from the solid on the fluid $\Qf$ 
in Eq.~\eqref{eq:JQFW2}. 
To capture the detail of the friction force and work, we examined the 
directional components of the total force vector $\bmFextsf$ in 
Eq.~\eqref{eq:frictionforce} in the Cartesian coordinate:
\begin{align}
\bmFextsf
&\equiv
\left(
\Fextsfx
    ,\ 
\Fextsfy
,\ 
\Fextsfz
   \right)^{\mathrm{T}}
\nonumber
\\
\Fextsfa
&=
\int_{z_{\text{SF}}}^{z_{\text{F}}} \rd z\:
    \rhof f_{\mathrm{sf},\alpha}^{\mathrm{ext}}
    \quad
    (\alpha = x,y,z)
    \label{eq:Fextsf_components}
\end{align}
and also decomposed the work 
$\Qf$ as the spatial integral of the inner product 
of the force density 
$\rho \bmfextsf=(
\rho {f}_{\mathrm{sf},x}^{\mathrm{ext}},\ 
\rho {f}_{\mathrm{sf},y}^{\mathrm{ext}},\ 
\rho {f}_{\mathrm{sf},z}^{\mathrm{ext}}
)^{\mathrm{T}}$ 
(see Eq.~\eqref{eq:def_fextsfx})
and velocity 
$\bmuf=({u}_{\mathrm{f},x}, {u}_{\mathrm{f},y}, {u}_{\mathrm{f},z})^{\mathrm{T}}$ 
into 
the three components as:
\begin{align}
\nonumber 
&\Qf =
\int_{z_{\text{SF}}}^{z_{\text{F}}} \rd z\:
    \rho \bmfextsf \cdot \bmuf 
=
\Qfxx + \Qfyy + \Qfzz,
\\
&\Qfaa =    
\int_{z_{\text{SF}}}^{z_{\text{F}}} \rd z\:
    \rho {f}_{\mathrm{sf},\alpha}^{\mathrm{ext}} {u}_{\mathrm{f},\alpha}
    (\alpha = x,y,z).
\label{eq:frictionwork}
\end{align}
%
Figure~\ref{fig:fig3} shows the directional 
components of the total force \add{per unit area} exerted on the liquid by the solid in 
Eq.~\eqref{eq:Fextsf_components}, and their contributions 
to the solid-liquid work as decomposed in Eq.~\eqref{eq:frictionwork}
for various wetting parameters $\eta$, with error bars representing 95~$\%$ confidence intervals, estimated from block averages over 5~ns.
%
At first for the force in Fig.~\ref{fig:fig3}(a), 
the total wall-normal force \add{per unit area} $\Fextsfz$ approximately corresponded to the 
control pressure of about 4~MPa, and $\Fextsfy$ was 
zero because the shear was not exerted in the $y$-direction. 
\add{Note that the resulting pressure slightly increased with wettability. 
This is because the top wall position was constrained before applying shear, 
while the heat generated by viscous dissipation increased with the velocity gradient, which increased with wettability.}
On the other hand, the absolute value of the shear stress $\Fextsfx$, corresponding 
to the integral of $\rho {f}_{\mathrm{sf},x}^{\mathrm{ext}}$ 
in Fig.~\ref{fig:fig2}(b), increased 
for larger wetting parameter $\eta$.
Regarding the error bars of the force, those for the 
wall-normal force 
$\Fextsfz$ were significantly larger than those for the 
wall-tangential ones of $\Fextsfx$ and $\Fextsfy$.
This is attributed to the roughness of the potential field formed by the solid walls. 
Thermodynamically, the fluctuations of the pressure in the normal and shear components are comparable; 
however, the instantaneous shear component of the solid-liquid force is determined by the instantaneous slip velocity and the FC. 
Thus, the magnitude of the fluctuation decreased as the FC decreased.
%
\par
\begin{figure}[bt]
  \begin{center}
    \includegraphics[width=.8\linewidth]{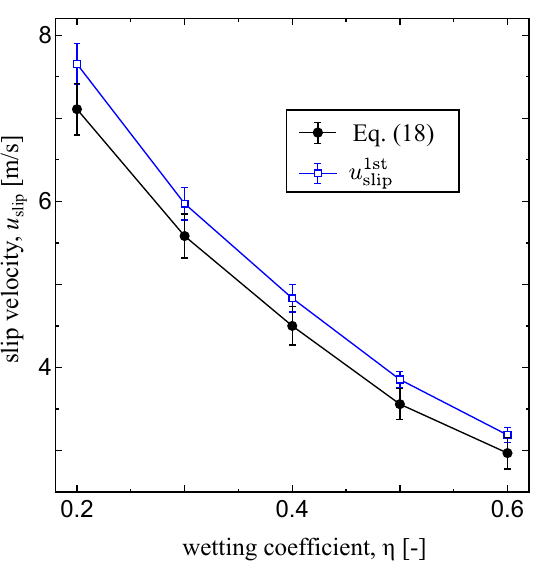}
  \end{center} 
  \caption{\label{fig:fig_uslip} 
  Slip velocity calculated using Eq.~\eqref{eq:u_slip1D} (shown in black circles) and that based on the average velocity of the first adsorption layer (shown in blue square) as a reference.
}
\end{figure}

With respect to the solid-liquid work in Fig.~\ref{fig:fig3}(b), 
both $\Qfyy$ and $\Qfzz$ were roughly equal to 
zero as expected for the present systems without flow in the 
$y$- and $z$-directions, although the error bars were relatively
large for $\Qfzz$ mainly because of the pressure fluctuation
as indicated in $\Fextsfz$.
%
%
\add{However, 
since the system is macroscopically one-dimensional, we can exclude this $\Qfzz$ term from the calculation of slip velocity by Eq.~\eqref{eq:u_slip1D}.}\rem{ Considering this large fluctuation, we excluded $\Qfzz$ 
from the calculation of slip velocity by Eq.~\eqref{eq:u_slip1D}}
\add{The results are shown as black circles }in Fig.~\ref{fig:fig2}\add{(c) and in} Fig.~\ref{fig:fig_uslip}\rem{ unwanted large 
errors due to thermal fluctuations}. 
Note that this treatment is applicable only for the present 
case where $\ufz=0$ can be assumed. 
In other words, $\Qfzz$ must be included in systems with non-zero wall-normal flow as 
exemplified in Fig.~\ref{fig:fig7} in Appendix~\ref{sec:AppA}.
%
%
%
\par
Finally, the slip velocity can be calculated by Eq.~\eqref{eq:u_slip1D} shown as black circles in Fig.~\ref{fig:fig_uslip} for various wettability parameters. 
With increasing the wetting parameter, the solid--fluid friction becomes larger and the slip velocity decreases. 
To compare our results with a conventional definition of the slip velocity, 
we also calculated the slip velocity based on the average velocity of the first adsorption layer $u_{\text{slip}}^{\text{1st}}$, which was defined as the integral of the fluid velocity weighted by the density of the first adsorption layer \add{and corresponded approximately to the velocity at the density peak}. 
Regardless of the wettability, the slip velocity evaluated by Eq.~\eqref{eq:u_slip1D} was smaller than that calculated based on the average velocity of the first adsorption layer. 
This supports again that our method was calculated by the integral of the fluid velocity weighted by the friction force, not by the density, \add{so that the resulting velocity was approximately the same as that at the peak position of the friction force.}
\add{Strictly speaking, the slip position cannot be uniquely defined in this study; however, in practice, 
the evaluated velocity corresponded well to that at the friction force peak. 
Therefore, the discrepancy in Fig.~\ref{fig:fig_uslip} can be attributed to the shift of} 
the peak position of the friction force\rem{ was slightly shifted} to the solid surface than that of the density. 
\begin{figure}[bt]
  \begin{center}
    \includegraphics[width=.8\linewidth]{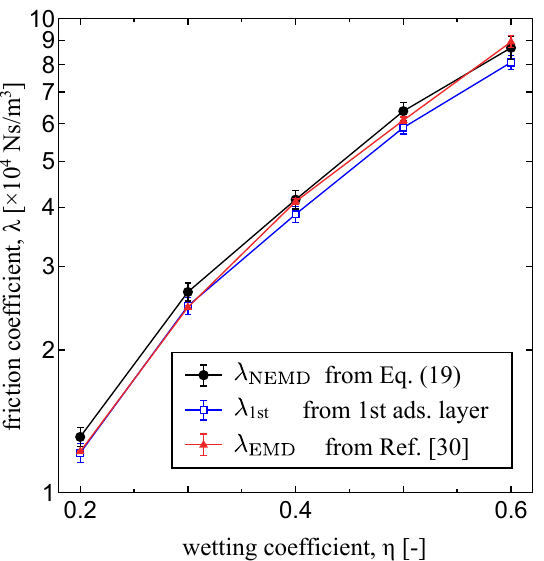}
  \end{center} 
  \caption{\label{fig:fig4} 
  Comparison of the friction coefficients calculated using Eq.~\eqref{eq:beta1D} (shown in black circles) and 
  Eq.~\eqref{eq:lambda1st} (shown in blue circles) and the Green-Kubo integral~\cite{oga_equilibrium_2023} (shown in red triangles) as a reference.
}
\end{figure}
\par
In addition, the friction coefficients (FC) can be evaluated by 
Eq.~\eqref{eq:beta1D} as black circles in Fig.~\ref{fig:fig4} for various wettability parameters, and 
we also calculated the FC based on the first \add{adsorption} layer as blue circles
\begin{equation}
    \lambda_{\text{1st}}=-\frac{\Fextsfx }{ u_{\text{slip}}^{\text{1D}}}.
    \label{eq:lambda1st}
\end{equation}
Owing to the discrepancy in the definition of the slip velocity, 
the FC based on the first adsorption layer was slightly underestimated compared to the present result. 
Figure~\ref{fig:fig4} also shows the EMD results based on the GK integral derived in a previous study~\cite{oga_equilibrium_2023} as a reference.
Note that the GK based method was derived considering the generalized Langevin equation and the one-dimensional momentum conservation, \ie the Stokes equation, and was used for EMD systems, 
whereas we considered the energy conservation at the interface in NEMD systems in this study.
\add{In other words, there is no guarantee that the two methods agree well because they use fundamentally different equations 
in the microscale system, where the boundary has non-zero thickness unlike macroscale. 
In addition, as often pointed out,~\cite{maffioli_slip_2022,oga_equilibrium_2023,Todd_Daivis2017,Evans2008} 
the GK based approach in EMD systems basically provides results in the zero-shear limit, whereas the NEMD approach is macroscopically at extremely high shear rate, 
and applying the recent method to overcome this problem~\cite{maffioli_slip_2022,MAFFIOLI2024109205} is a possible choice for further comparison between the EMD and NEMD results. However,}
in spite of the differences in the methodology and the system, the present results 
agreed well with the GK results especially for large wettability of $\eta > 0.4$. 
\par
Regarding this trend, Fig.~\ref{fig:fig2}(c) implies that it is attributed to the shift in the boundary position.
Specifically for the nanoscale SL interface, the boundary based on the energy conservation may approach the solid surface with lower wettability whereas the boundary based on the momentum conservation approaches the solid surface with higher wettability~\cite{herrero_shear_2019}. 
Therefore, for the lower wettability, the GK based results agreed well with the FC based on the first adsorption layer, implying that the boundary position based on the momentum conservation varies between the peaks of the fluid density and the friction force near the solid surface.
This discrepancy may be an intrinsic issue that inevitably arises when representing a nanoscale interface with non-zero thickness as a macroscopic boundary with zero thickness.
Therefore, at this stage, it still remains unclear which method provides the correct friction coefficient, or there can be multiple possibilities for the definition of the interface position.
\add{It is worth noting that 
our method successfully avoids the arbitrariness in the definition of the slip velocity and friction coefficient, 
and enables the extraction of the macroscopic boundary of the heat transport at the solid surface via the molecular simulation; however, the boundary position has not yet been clarified in this study.}
\par
\add{Moreover, our method was derived for a macroscopically steady-state one-dimensional system and has been validated only for a Lennard-Jones fluid. 
Therefore, it cannot be directly applied to extract the spatial distribution of the slip velocity, \eg near dynamic contact lines.
From a theoretical viewpoint, our method also requires a fluid region sufficiently far from the solid wall, as well as the calculation of the friction force distribution. 
Consequently, further extension is needed, for instance, for very narrow channels, where the fluid molecules directly interact with both walls, 
or for systems with friction due to long-range solid--fluid interaction, where the fluid--solid interaction force cannot be independently defined due to its long-range nature.
}

The methodology derived in this study can be extended to a system with a nonflat wall surface 
and enables the calculation of slip velocity without defining the slip boundary position, unlike in previous studies. 
However, the results obtained from the present method may not necessarily agree with the GK based results. 
This is because our formulation is based only on energy conservation, 
whereas the GK based method is based on the generalized Langevin equation of solid wall and the one-dimensional Stokes equation of the fluid, 
which makes the boundary treatment unclear and 
considers the interaction between 
bulk-like liquid and solid-like regions. 
In other words, the GK based method was intended to represent an interface --- even one with nanoscale structure ---
as macroscopically smooth using the Stokes equation, 
which raised an ambiguous definition of the boundary. 
On the other hand, the direct NEMD method in this study 
represents the interfacial structures as two-dimensional, 
which raises concerns upon addressing the macroscopic boundaries. 
This discrepancy becomes clearer at the interface with nanoscale structure. 
Specifically, 
\add{in Appendix~\ref{sec:AppA}, }
we also conducted calculations for a system with a grooved surface, which increased the ambiguity of the boundary
\add{and resulted in the discrepancy of the FC from the GK based method. 
See Appendix~\ref{sec:AppA} for further details. 
Such a discrepancy is attributed to the use of fundamentally different equations and the associated differences in the treatment of the boundary. 
Evaluating the spatial variations of friction force and velocity fields in the first adsorption layer formed around nanoscale structures is an interesting future topic.}
%
\par
\par
\section*{Conclusion}
In this study, we developed a methodology for extracting slip velocity in non-equilibrium molecular dynamics (NEMD) simulations based on a thermal perspective from macroscopic and microscopic viewpoints. 
At a solid--fluid interface under shear, 
heat was generated owing to frictional dissipation, which induced a discontinuous heat flux there. 
At the macroscale, frictional heat was defined as the product of the friction force and slip velocity, 
whereas at the microscale, 
it can be expressed as the sum of the works exerted on the fluid and solid by each other.
By combining the two different scales, the slip velocity 
was obtained by the integration of the fluid velocity weighted by the solid--fluid friction force. 
As a test calculation, we conducted NEMD simulations of a quasi-one-dimensional Couette-type flow system of a Lennard--Jones liquid, 
and we found that the resulting slip velocity corresponded to the relative velocity of the fluid closer to the solid surface than both the fluid density peak and the average velocity of the first adsorption layer.
We also calculated the friction coefficient and compared with the Green-Kubo integral obtained by EMD simulations, 
and found that they agreed well with high wettability. 
In contrast, with the low wettability, they showed a slight discrepancy and can be attributed to the different frameworks. 
%
%
%

\begin{acknowledgements}
We thank Haruki Oga for fruitful discussion as a former member of Y.Y.’s group at the University of Osaka.
H.K. was supported by JSPS KAKENHI Grant No. JP23KJ0090, and Y.Y. was supported by
JSPS KAKENHI Grant Nos.  JP22H01400 and JP23H01346. 
Numerical simulations were performed on the Supercomputer system"AFI-NITY II" at the Advanced Fluid Information Research Center, Institute of Fluid Science, Tohoku University.
\end{acknowledgements}
\appendix
\section{Slip velocity at a grooved surface}
\label{sec:AppA}
\begin{figure}[tb]
  \begin{center}
    \includegraphics[width=.9\linewidth]{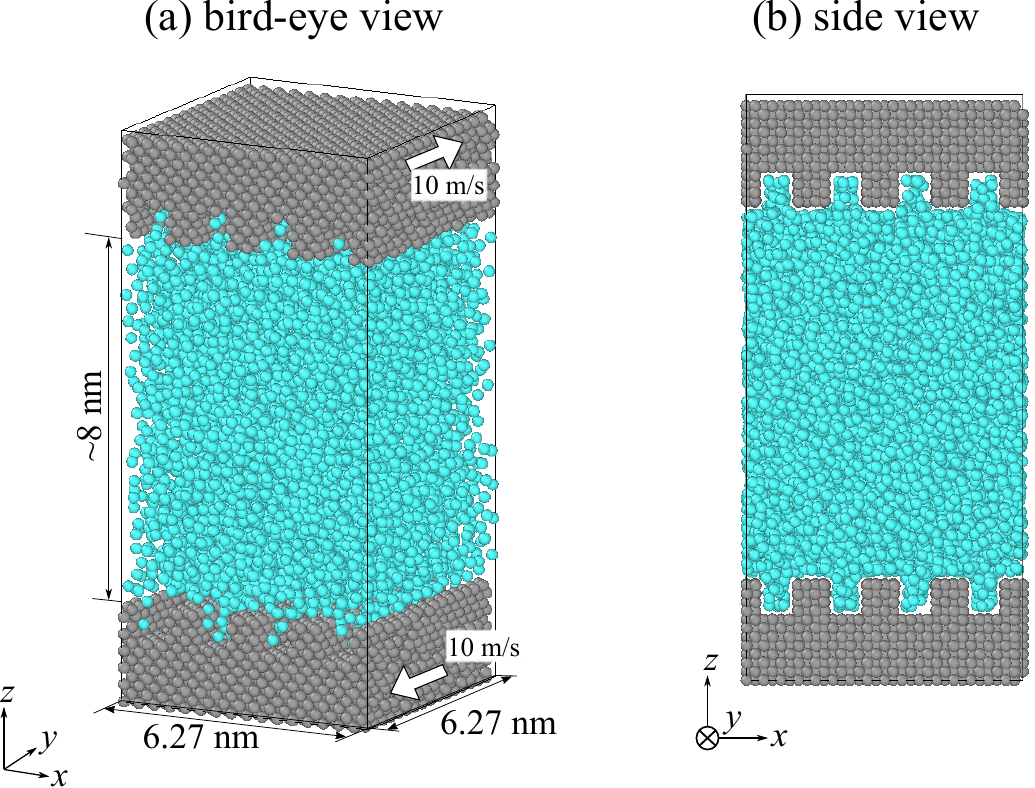}
  \end{center} 
  \caption{\label{fig:fig7} 
  (a) Bird-eye view and (b) side view of the system with a grooved surface.
}
\end{figure}

Non-equilibrium molecular dynamics (NEMD) simulations were performed to analyze the solid-fluid friction of a grooved surface as shown in Fig.~\ref{fig:fig7}. 
The basic setup was similar to that in Fig.~\ref{fig:fig1} 
whereas the wetting parameters were chosen as  $\eta=0.1-0.3$ and 
the wall velocity was applied in the $y$-direction parallel to the groove.

\begin{figure}[tb]
  \begin{center}
    \includegraphics[width=.9\linewidth]{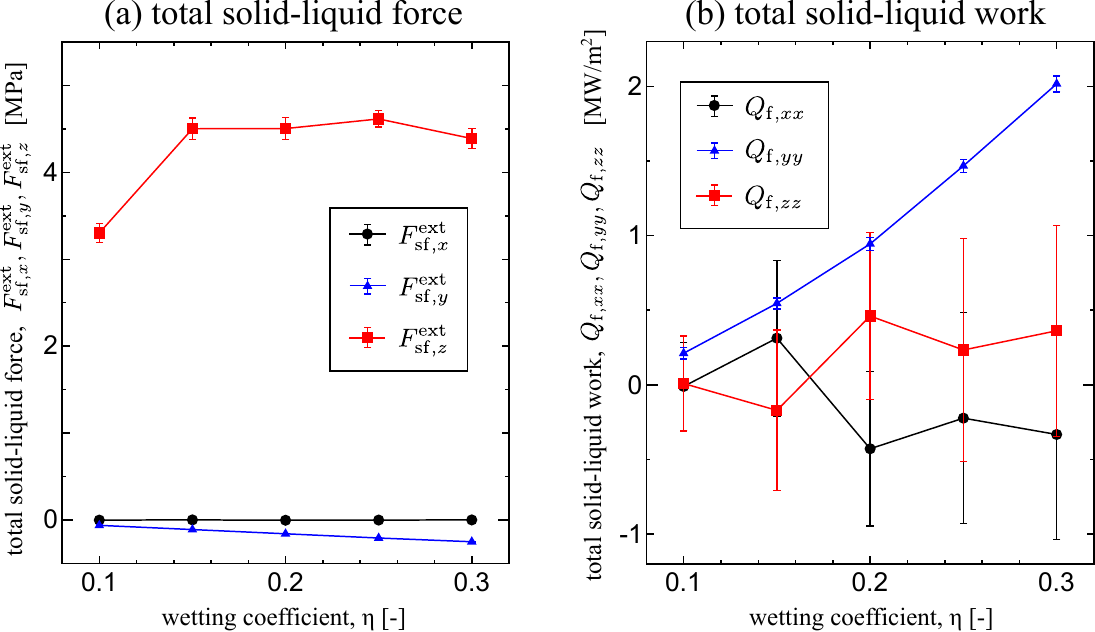}
  \end{center} 
  \caption{\label{fig:fig8} 
  Three components of the integrals of the (a) friction force and (b) friction work per unit area exerted on the fluid by the wall for various wetting parameters.
}
\end{figure}
Figure~\ref{fig:fig8}(a) and (b) show the friction force per unit area obtained using Eq.~\eqref{eq:frictionforce} and the friction work per unit area obtained using Eq.~\eqref{eq:frictionwork},
respectively.
Although the resulting pressure was approximately equal to the control pressure of 4~MPa, similar to the results in Fig.~\ref{fig:fig3}, 
the integral of the friction work in the $x$-direction had a larger error than that in Fig.~\ref{fig:fig3}. 
Moreover, 
the amplitude of the error was comparable to that in the $z$-direction. 
In this system, no macroscopic flow was expected in the $x$- and $z$-directions, and only the $y$-direction friction work was used to calculate the slip velocity and friction coefficient (FC).
Owing to this error in the wall-normal direction, when applying the wall velocity in the $x$-direction, the calculation of the slip velocity and FC is difficult.

\begin{figure}[t]
  \begin{center}
    \includegraphics[width=.7\linewidth]{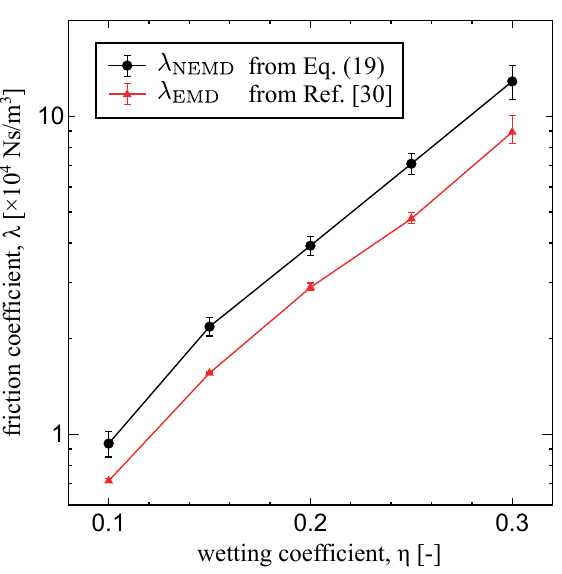}
  \end{center} 
  \caption{\label{fig:fig9} 
      Friction coefficients calculated using Eq.~\eqref{eq:beta1D} (shown in black circles) and the Green-Kubo integral~\cite{oga_equilibrium_2023} (shown in red triangles) as a reference.
}
\end{figure}

Based on the results in Fig.~\ref{fig:fig8}, Fig.~\ref{fig:fig9} shows the FC in the $y$-direction of the grooved surface 
compared with the results based on the Green-Kubo method in Ref.~\citenum{oga_equilibrium_2023}. 
The results by the two methods were apparently different; however, it is understandable 
because the method derived in this study was based on the frictional dissipation between the solid and fluid whereas the Green-Kubo method in Ref.~\citenum{oga_equilibrium_2023} was basically based on one-dimensional Stokes equation assuming flat surface.


%
%
\section*{Conflict of Interest Statement }
The authors have no conflicts to disclose.

\section*{Data Availability Statement}

The data that support the findings of this study are available from the corresponding author upon reasonable request.
\bibliography{ref}

%

\end{document}
%